# Algebraic theory of linear viscoelastic nemattodynamics
# Part 2: Linear viscoelastic nematic viscoelasticity

Arkady I. Leonov

*Department of Polymer Engineering, The University of Akron,*

*Akron, Ohio 44325-0301, USA.*

**Abstract**

This *second part* of paper develops a theory of linear viscoelastic nematodynamics applicable to LCP. The viscous and elastic nematic components in theory are described by using the LEP approach for viscous nematics and de Gennes free energy for weakly elastic nematic elastomers. The case of applied external magnetic field exemplifies the occurrence of non-symmetric stresses. In spite of multi- (10) parametric character of the theory, the use of nematic operators presents it in an elegant form. When the magnetic field is absent, the theory is simplified for symmetric case with 6 parameters, and takes an extremely simple, 2-parametric form for viscoelastic nematodynamics with possible *soft deformation modes*. It is shown that the linear nematodynamics is always reduced to the LEP-like equations where the coefficients are changed for linear memory functionals whose parameters are calculated from original viscosities and moduli.

## Introduction

It is now well recognized that the continual theories of viscous nematodynamics and nematic elasticity describe well the behavior of liquid crystals (LC's) (e.g. see the texts [1,2] and cited papers) and liquid crystalline elastomers (LCE's) (e.g. see the text [3]. and cited papers), respectively. Recent publications [4,5] also showed that the well-known Leslie-Ericksen-Parodi (LEP) continual theory of nematic LCs has a complete continual analog for description of weakly nonlinear elastic behavior of LCE's, when using the de Gennes elastic potential. Also, the papers [4,5] rigorously analyzed the possible soft deformation nematic modes in both the elastic and viscous cases.

The respective theories for liquid crystalline polymers (LCP's) and LCE's are not that well developed (see, however, the recent dynamic theory [3,6] for LCE's). In the most difficult case of LCP, the most developed is the Doi molecular "rigid rod" approach (e.g. see the text [7]) with further elaborations discussed in [2]. As an alternative to the Doi theory, the Poisson-Bracket continuum approach was applied for developing constitutive equations LCP's (e.g. see the text [8]). This theory reduces to the Doi theory in the homogeneous (monodomain) limit. All the abovementioned theories for LCP employ the same state variables as in case of LMW liquid crystals, i.e. the director $\underline{n}$ (or the second rank order tensor), and the director's space gradient $\underline{\nabla n}$.

To take into account the macromolecular flexibility, which along with rigidity of LCP macromolecules, plays also important role in their dynamics, several attempts have been made to develop viscoelastic nematodynamics. Paper [9] made *ad hoc* viscoelastic extension of LEP equation. This extension was used later [6] in attempt to involve molecular structure in this type of LCP description. Continuum approaches, not related to thermodynamics [10,11], and linear theories related to thermodynamics [12-15], have been also proposed, although the approach of papers [14,15] is conceptually doubtful.

It seems that the awkward common tensor/matrix form of operations in nematic theories does not allow displaying their simple algebraic structure whose ignorance makes difficult (if possible) elaborating a general theory of linear or weakly nonlinear dynamic behavior of LCP's. This paper reveals the algebraic structure of the nematic theories and presents it in a simple form. This allows us to demonstrate remarkable features of linear nematic viscoealsticity of LCP.

## 2.1. Linear kinematical relations

Bearing in mind that the main objective of this part is elaborating nematic viscoelasticity, we avoid discussing here the well-known general relations for balance of moment of momentum and related rotational inertia effects. These relations are discussed in monographs by de Gennes and Prost [1], Warner and Terentjev [3] and in more detail, in work by Leonov and Volkov [16].

We use in the following the linear viscoelastic kinematics based on the common decomposition of full infinitesimal strain gradient tensor $\underline{\underline{\gamma}}$ into elastic (transient) $\underline{\underline{\gamma}}_e$ and inelastic $\underline{\underline{\gamma}}_p$ parts. Each of these tensors consists of strain and rotation tensors, i.e. $\left(\underline{\underline{\gamma}},\underline{\underline{\gamma}}_e,\underline{\underline{\gamma}}_p\right)=\left(\underline{\underline{\varepsilon}},\underline{\underline{\varepsilon}}_e,\underline{\underline{\varepsilon}}_p\right)+\left(\underline{\underline{\Omega}}^b,\underline{\underline{\Omega}}_e^b,\underline{\underline{\Omega}}_p^b\right)$, where $\left(\underline{\underline{\varepsilon}},\underline{\underline{\varepsilon}}_e,\underline{\underline{\varepsilon}}_p\right)$ are the infinitesimal strain tensors and $\left(\underline{\underline{\Omega}}^b,\underline{\underline{\Omega}}_e^b,\underline{\underline{\Omega}}_p^b\right)$ are infinitesimal "body" rotations. So the decomposition is presented as $\underline{\underline{\gamma}}=\underline{\underline{\gamma}}_e+\underline{\underline{\gamma}}_p$. Differentiating this equation with respect to time and separating equations for deformation and rotations, yields the rate equations:

$$\dot{\underline{\underline{\varepsilon}}}_e+\underline{\underline{e}}_p=\underline{\underline{e}} \tag{2.1}$$

$$\dot{\underline{\underline{\Omega}}}_e^b+\underline{\underline{\omega}}_p^b=\underline{\underline{\omega}}^b \tag{2.2$_1$}$$

Here overdots denote time derivatives, $\underline{\underline{e}}=1/2[\underline{\underline{\nabla \mathrm{v}}}+(\underline{\underline{\nabla \mathrm{v}}})^T]$ and $\underline{\underline{\omega}}^b=1/2[\underline{\underline{\nabla \mathrm{v}}}-(\underline{\underline{\nabla \mathrm{v}}})^T]$ are, respectfully, the full strain rate and vorticity of "body", and $\underline{\mathrm{v}}$ is the velocity vector. So $\dot{\underline{\underline{\gamma}}}\equiv\underline{\underline{\nabla \mathrm{v}}}=\underline{\underline{e}}+\underline{\underline{\omega}}^b$, $\underline{\underline{e}}_p=\dot{\underline{\underline{\varepsilon}}}_p$ and $\underline{\underline{\omega}}_p^b=\dot{\underline{\underline{\Omega}}}_p^b$ are, respectfully, the irreversible strain rate and vorticity.

We now introduce in (2.1), (2.2$_1$) the additional rate equation describing *internal rotations* for nematic continuum:

$$\dot{\underline{\underline{\Omega}}}_e^i+\underline{\underline{\omega}}_p^i=\underline{\underline{\omega}}^i. \tag{2.2$_2$}$$

Here, following Pleiner and Brand [12,13] we decomposed the total tensor of internal vorticity $\underline{\underline{\omega}}^i$ into the rate of reversible rotation $\dot{\underline{\underline{\Omega}}}_e^i$ and irreversible vorticity $\underline{\underline{\omega}}_p^i$. Extracting now ($2.2_2$) from ($2.2_1$) results in the final equation describing the linear kinematics of *relative rotations* in nematic continuum:

$$\dot{\underline{\underline{\Omega}}}_e + \underline{\underline{\omega}}_p = \underline{\underline{\omega}} \qquad \left(\underline{\underline{\Omega}}_e = \underline{\underline{\Omega}}_e^b - \underline{\underline{\Omega}}_e^i,\ \underline{\underline{\omega}}_p = \underline{\underline{\omega}}_p^b - \underline{\underline{\omega}}_p^i,\ \underline{\underline{\omega}} = \underline{\underline{\omega}}^b - \underline{\underline{\omega}}^i\right). \tag{2.2$_3$}$$

The evolution of director, generally characterized by kinematical equation, $\dot{\underline{n}} = \underline{\omega}^i \times \underline{n} = -\underline{\underline{\omega}}^i \cdot \underline{n}$, in the linear nematic theory near the state with a constant value of director, $\underline{n} = const$, can be considered as the evolution of director disturbance as:

$$d(\delta \underline{n})/dt = -\underline{\underline{\omega}}^i \cdot \underline{n}\,. \tag{2.3}$$

We finally introduce the kinematical variables, convenient for characterizing a combined effect of viscoelastic deformations and relative rotations:

$$\overset{0}{\underline{\underline{\Gamma}}}_e + \underline{\underline{\gamma}}_p = \underline{\underline{\gamma}} \qquad \left(\underline{\underline{\Gamma}}_e = \underline{\underline{\varepsilon}} + \underline{\underline{\Omega}}_e,\ \underline{\underline{\gamma}}_p = \underline{\underline{e}}_p + \underline{\underline{\omega}}_p,\ \underline{\underline{\gamma}} = \underline{\underline{e}} + \underline{\underline{\omega}}\right). \tag{2.4}$$

We consider below the simple, incompressible case, when the all components of strain and their rate are *traceless*.

## 2.2. Thermodynamics and constitutive relations

### *2.2.1. Non-symmetric theory*

To describe the *quasi-equilibrium effects* in weak nematic viscoelasticity we will use the de Gennes type of potential (Helmholtz free energy density):

$$f = 1/2G_0 \left|\underline{\underline{\varepsilon}}\right|^2 + G_1 \underline{n}\underline{n} : \underline{\underline{\varepsilon}}^2 + G_2 (\underline{n}\underline{n} : \underline{\underline{\varepsilon}})^2 - 2G_3 \underline{n}\underline{n} : (\underline{\underline{\varepsilon}} \cdot \underline{\underline{\Omega}}_e) - G_5 \underline{n}\underline{n} : \underline{\underline{\Omega}}_e^2 \tag{2.5}$$

Here $G_k$ are the elastic moduli, and the lower index in $\underline{\underline{\varepsilon}}_e$ is missed for simplicity.

Using now the expression for the dissipation $D$ in the system,

$$D \equiv TP_s\big|_T = \underline{\underline{\sigma}}^s : \underline{\underline{e}} + \underline{\underline{\sigma}}^a : \underline{\underline{\omega}} - df\big|_T / dt,$$

represents it with the aid of (2.5) in the form:

$$D = \underline{\underline{\sigma}}_p^s : \underline{\underline{e}} + \underline{\underline{\sigma}}_p^a : \underline{\underline{\omega}} + \underline{\underline{\sigma}}_e^s : \underline{\underline{e}}_p + \underline{\underline{\sigma}}_e^a : \underline{\underline{\omega}}_p\,.$$

Here

$$\underline{\underline{\sigma}}_p^s \equiv \underline{\underline{\sigma}}^s - \underline{\underline{\sigma}}_e^s, \quad \underline{\underline{\sigma}}_p^a \equiv \underline{\underline{\sigma}}^a - \underline{\underline{\sigma}}_e^a, \quad \underline{\underline{\sigma}}_e^s = \partial f / \partial \underline{\underline{\varepsilon}}, \quad \underline{\underline{\sigma}}_e^a = \partial f / \partial \underline{\underline{\Omega}}_e . \tag{2.6}$$

In (2.6), $\underline{\underline{\sigma}}_e^s$ ($\underline{\underline{\sigma}}_p^s$) and $\underline{\underline{\sigma}}_e^a$ ($\underline{\underline{\sigma}}_p^a$) are the equilibrium (non-equilibrium) symmetric and asymmetric parts of the extra stress. Additionally, $T$ is the temperature, $P_s$ is the entropy production, $\underline{\underline{\sigma}}^s$ and $\underline{\underline{\sigma}}^a$ are the symmetric and asymmetric parts of the extra stress tensor, respectively.

Keeping in mind possible applications to LCP we consider below only the simplest (Maxwell-like) liquid case $\underline{\underline{\sigma}}_p^s = 0, \quad \underline{\underline{\sigma}}_p^a = 0$ with the dissipation presented as:

$$D = \underline{\underline{\sigma}}^s : \underline{\underline{e}}_p + \underline{\underline{\sigma}}^a : \underline{\underline{\omega}}_p . \tag{2.7}$$

Due to (2.5) the symmetric $\underline{\underline{\sigma}}^s$ and asymmetric $\underline{\underline{\sigma}}_h^a$ parts of extra stress traceless tensor are defined as:

$$\begin{aligned} \underline{\underline{\sigma}}^s = \partial f / \partial \underline{\underline{\varepsilon}} = G_0 \underline{\underline{\varepsilon}} + G_1 [\underline{n}\underline{n} \cdot \underline{\underline{\varepsilon}} + \underline{\underline{\varepsilon}} \cdot \underline{n}\underline{n} - 2nn(\underline{\underline{\varepsilon}} : \underline{n}\underline{n})] + 2(G_1 + G_2)(\underline{n}\underline{n} - \underline{\underline{\delta}}/3)(\underline{\underline{\varepsilon}} : \underline{n}\underline{n}) \\ + G_3 (\underline{n}\underline{n} \cdot \underline{\underline{\Omega}}_e - \underline{\underline{\Omega}}_e \cdot \underline{n}\underline{n}), \end{aligned} \tag{2.8$_1$}$$

$$\underline{\underline{\sigma}}^a = \partial f / \partial \underline{\underline{\Omega}}_e = -G_4 (\underline{n}\underline{n} \cdot \underline{\underline{\varepsilon}} - \underline{\underline{\varepsilon}} \cdot \underline{n}\underline{n}) + G_5 (\underline{n}\underline{n} \cdot \underline{\underline{\Omega}}_e + \underline{\underline{\Omega}}_e \cdot \underline{n}\underline{n}). \tag{2.8$_2$}$$

Here the Onsager relation $G_4 \equiv -G_3$ automatically follows from (2.5).

Among several sources of stress asymmetry, such as inertial effects of internal rotations, orientation (Frank) elasticity and the Cosserat/Born isotropic couples, the most important for LCP is the action of external magnetic field, $\underline{H}$. Under the common assumptions, the magnetic field is potential, with the potential function $\Psi = -1/2 \underline{\underline{\chi}}(\underline{n}) : \underline{H}\underline{H}$. Here $\underline{\underline{\chi}}(\underline{n}) = \chi_{\perp} \underline{\underline{\delta}} + \chi_a \underline{n}\underline{n}$ is the susceptibility tensor, $\chi_{\|}$ and $\chi_{\|}$ are the susceptibilities parallel and perpendicular to the director, with $\chi_a = \chi_{\|} - \chi_{\perp}$ being the magnetic anisotropy. Because of a typical weak magnetization of the considered diamagnetic liquid, the effect of magnetic field on constitutive parameters $G_k$ in (2.5) can be neglected. In this case, the body couple ((“effective magnetic field”) is defined as:

$$\underline{h} = -\partial \Psi / \partial \underline{n} = \chi_a \underline{H} (\underline{n} \cdot \underline{H}) . \tag{2.9}$$

where $\chi_a$ is the magnetic anisotropy. When the inertial effects of internal rotations are negligible, the equilibrium equation for internal torques in magnetic field is:

$$\underline{\underline{\sigma}}^a = 1/2(\underline{h}\underline{n} - \underline{n}\underline{h}), \quad \underline{\underline{\sigma}}^a \cdot \underline{n} = 1/2[\underline{h} - \underline{n}(\underline{h}\cdot\underline{n})] = 1/2\chi_a(\underline{n}\cdot\underline{H})[\underline{H} - \underline{n}(\underline{n}\cdot\underline{H})]. \quad (2.10)$$

The relation (2.10) shows that *in the absence of magnetic field, stress tensor is symmetric*.

Due to (2.7), the constitutive relations between the irreversible kinematical variables and stress are of the LEP type:

$$\begin{aligned}\underline{\underline{\sigma}}^s &= \eta_0 \underline{\underline{e}}_p + \eta_1[\underline{nn}\cdot\underline{\underline{e}}_p + \underline{\underline{e}}_p\cdot\underline{nn} - 2\underline{nn}(\underline{nn}:\underline{\underline{e}}_p)] + 2(\eta_1+\eta_2)(\underline{nn} - \underline{\underline{\delta}}/3)(\underline{nn}:\underline{\underline{e}}_p) \\ &\quad + \eta_3(\underline{nn}\cdot\underline{\underline{\omega}}_p - \underline{\underline{\omega}}_p\cdot\underline{nn})\end{aligned}, \quad (2.11_1)$$

$$\underline{\underline{\sigma}}^a = -\eta_4(\underline{nn}\cdot\underline{\underline{e}}_p - \underline{\underline{e}}_p\cdot\underline{nn}) + \eta_5(\underline{nn}\cdot\underline{\underline{\omega}}_p + \underline{\underline{\omega}}_p\cdot\underline{nn}) \quad (\eta_4 = -\eta_3). \quad (2.11_2)$$

Here we used the Onsager relation: $\eta_4 = -\eta_3$. Under common assumptions, we consider the kinetic coefficients $\eta_k$, as well as the parameters $G_k$, being independent of $\underline{H}$. To avoid the degeneration of CE's we further assume that $G_k \neq 0$ and $\eta_k \neq 0$.

Demanding the elastic potential (2.5) to be thermodynamically stable results in the necessary and sufficient stability conditions [4], imposed on the parameters $G_k$:

$$G_0 > 0;\ \ G_5 > 0;\ \ G_0 + G_1 > 0;\ \ 3/4G_0 + G_1 + G_2 > 0;\ \ (G_0+G_1)G_5 > G_3^2. \quad (2.12_1)$$

The thermodynamic stability conditions for dissipation in (2.8) in incompressible case are the same as in (2.12$_1$) with substitution $G_k \to \eta_k$, i.e. [5]:

$$\eta_0 > 0;\ \ \eta_5 > 0;\ \ \eta_0 + \eta_1 > 0;\ 3/4\eta_0 + \eta_1 + \eta_2 > 0;\ \ (\eta_0+\eta_1)\eta_5 > \eta_3^2 \quad (2.12_2)$$

Substituting now CE's (2.9$_{1,2}$) and (2.10) into the expression for dissipation (2.7) reduces it to the quadratic form:

$$D = \eta_0\left|\underline{\underline{e}}_p\right|^2 + 2\eta_1\underline{nn}:\underline{\underline{e}}_p^2 + 2\eta_2(\underline{nn}:\underline{\underline{e}}_p)^2 - 4\eta_3\underline{nn}:(\underline{\underline{e}}_p\cdot\underline{\underline{\omega}}_p) - 2\eta_5\underline{nn}:\underline{\underline{\omega}}_p^2, \quad (2.13)$$

which due to the stability constraints (2.12$_{1\text{-}3}$) is positively definite.

The variables $\underline{\underline{\omega}}_p$ and $\underline{\underline{e}}_p$ will be further excluded from the final formulation, giving place to coupled equations for evolution of the hidden thermodynamic parameters $\underline{\underline{\Omega}}_e$ and $\underline{\underline{\varepsilon}}$.

We remind once again that in the absence of magnetic field ($\underline{H} = 0$), $\underline{\Sigma} = \underline{\underline{\sigma}}^a\cdot\underline{n} = 0$, so stress tensor is symmetric.

### 2.2.2. *Symmetric theory*

In symmetric case when $\underline{\underline{\sigma}}^a = 0$, equations ($8_2$), ($9_2$) yield the kinematical relations:

$$\underline{\underline{\Omega}}_e = \lambda_1(\underline{\underline{\varepsilon}}\cdot\underline{nn} - \underline{nn}\cdot\underline{\underline{\varepsilon}}), \quad \underline{\underline{\omega}}_p = \lambda_2(\underline{\underline{e}}_p\cdot\underline{nn} - \underline{nn}\cdot\underline{\underline{e}}_p)$$
$$(\lambda_1 = G_3/G_5,\ \lambda_2 = \eta_3/\eta_5) \qquad (2.14_{1,2})$$

where $\lambda_1$ and $\lambda_2$ are sign indefinite parameters. Substituting ($2.14_{1,2}$) back into ($8_2$), ($9_2$), and in (5) yields the reduced (or "renormalized") formulation of these equations with symmetric stress as:

$$f = 1/2G_0\left|\underline{\underline{\varepsilon}}\right|^2 + G_1^r\,\underline{nn}:\underline{\underline{\varepsilon}}^2 + G_2^r(\underline{nn}:\underline{\underline{\varepsilon}})^2 \quad (G_1^r = G_1 - G_3^2/G_5,\ G_2^r = G_2 + G_3^2/G_5) \quad (2.15)$$

$$\underline{\underline{\sigma}} = \partial f/\partial\underline{\underline{\varepsilon}} = G_0\underline{\underline{\varepsilon}} + G_1^r[\underline{nn}\cdot\underline{\underline{\varepsilon}} + \underline{\underline{\varepsilon}}\cdot\underline{nn} - 2\underline{nn}(\underline{\underline{\varepsilon}}:\underline{nn})] + 2(G_1^r + G_2^r)(\underline{\underline{\varepsilon}}:\underline{nn})(\underline{nn} - 1/3\underline{\underline{\delta}}) \quad (2.16_1)$$

$$\underline{\underline{\sigma}} = \frac{1}{2}\frac{\partial D}{\partial \underline{\underline{e}}_p} = \eta_0\underline{\underline{e}}_p + \eta_1^r[\underline{nn}\cdot\underline{\underline{e}}_p + \underline{\underline{e}}_p\cdot\underline{nn} - 2\underline{nn}(\underline{\underline{e}}_p:\underline{nn})] + 2(\eta_1^r + \eta_2^r)(\underline{\underline{e}}_p:\underline{nn})(\underline{nn} - \underline{\underline{\delta}}/3) \quad (2.16_2)$$

Additionally, the dissipation $D = TP_s\big|_T = \underline{\underline{\sigma}}:\underline{\underline{e}}_p$ is presented in the form:

$$D/2 = \eta_0\left|\underline{\underline{e}}_p\right|^2 + \eta_1^r\,\underline{nn}:\underline{\underline{e}}_p^2 + \eta_2^r(\underline{nn}:\underline{\underline{e}}_p)^2 \quad (\eta_1^r = \eta_1 - \eta_3^2/\eta_5,\ \eta_2^r = \eta_2 + \eta_3^2/\eta_5) \quad (2.17)$$

Equation ($2.16_1$) is the Ericksen CE. It is seen that D/2 there plays the role of Raleigh (potential) dissipative function.

The (necessary and sufficient) conditions of *thermodynamic stability* ($1.11_{1,2}$) are:

$$G_0 > 0,\ G_0 + G_1^r = G_0 + G_1 - G_3^2/G_5 > 0,\ 3/4G_0 + G_1^r + G_2^r = 3/4G_0 + G_1 + G_2 > 0 \quad (2.18_1)$$

$$\eta_0 > 0, \quad \eta_0 + \eta_1^r = \eta_0 + \eta_1 - \eta_3^2/\eta_5 > 0, \quad 3/4\eta_0 + \eta_1^r + \eta_2^r = 3/4\eta_0 + \eta_1 + \eta_2 > 0. \quad (2.18_2)$$

One can see the complete similarity in inequalities ($2.18_{1,2}$). For simplicity we change hereafter the notations as: $G_k^r \leftrightarrow G_k$, $\eta_k^r \leftrightarrow \eta_k$ $(k = 1, 2)$.

## 2.3. N-operators in viscoelastic nematodynamics

The following correspondence exists between the continual equations in Section 2.2.and algebraic relations in Section 1.1, shown in Table 2.

*Table2. Correspondence between algebraic and physical variables/parameters/equations*

| Algebraic | Elastic Nematic | Viscous Nematic |
|---|---|---|
| $\underline{\underline{x}}_s$ ……………. ….(1.1)<br>symmetric tensor (variable) | $\underline{\underline{\varepsilon}}$ …..…………..…(2.8$_{1,2}$)<br>elastic strain tensor | $\underline{\underline{e}}_p$ …………….. (2.11$_{1,2}$)<br>inelastic strain rate tensor |
| $\underline{\underline{x}}_a$ ………………..(1.1)<br>asymmetric tensor variable | $\underline{\underline{\Omega}}_e$ ………………(2.8$_{1,2}$)<br>elastic relative rotation | $\underline{\underline{\omega}}_p$ ……………..(2.11$_{1,2}$)<br>inelastic relative vorticity |
| $\underline{\underline{y}}_s$ ……………….(1.1)<br>symmetric tensor, function | $\underline{\underline{\sigma}}^s$ ………………(2.8$_{1,2}$)<br>symmetric extra stress | $\underline{\underline{\sigma}}^s$ ……………..(2.11$_{1,2}$)<br>asymmetric extra stress |
| $\underline{\underline{y}}_a$ ……………….(1.1)<br>asymmetric tensor, function | $\underline{\underline{\sigma}}^a$ ………………(2.8$_{1,2}$)<br>symmetric extra stress | $\underline{\underline{\sigma}}^a$ ……………(2.11$_{1,2}$)<br>asymmetric extra stress |
| $r_k$ $(k=0,..,5)$ ………(1.1)<br>parameters of N-operators | $\hat{G}_k$ … ……..(2.8$_{1,2}$, 2.14$_1$)<br>nematic elastic moduli | $\hat{\eta}_k$ … (2.8$_{1,2}$, 2.14$_2$)<br>nematic viscosities |
| $P$ …………………(1.2)<br>quadratic form | $f$ ……………… (2.5)<br>nematic free energy | $D$ ……………….(2.13)<br>dissipation |

The constitutive parameters $\hat{G}_k$ and $\hat{\eta}_k$ in Table 2 are:

$$\hat{G}_k = G_k \;\; (k=0,1,3,4,5), \;\; \hat{G}_2 = 2G_1 + 2G_2, \;\; G_4 = -G_3 \qquad (2.19_1)$$

$$\hat{\eta}_k = \eta_k \;\; (k=0,1,3,4,5), \;\; \hat{\eta}_2 = 2\eta_1 + 2\eta_2, \;\; \eta_4 = -\eta_3\,. \qquad (2.19_2)$$

Non-degenerating conditions $G_k \neq 0, \;\; \eta_k \neq 0$ are considered below. Using Table 2 it is easy establish direct and inverse relations for CE's described by ON-operators.

### *2.3.1 N-operator presentation of non-symmetric theory*

1) *N-operator presentation of "elastic" and "viscous" CE's* (2.8$_{1,2}$) *and* (2.9$_{1,2}$):

$$\underline{\underline{\sigma}} \equiv \mathbf{G}(\underline{n}) \bullet \underline{\underline{\Gamma}}_e = \sum_{k=0}^{5} \hat{G}_k \mathbf{a}_k(\underline{n}) \bullet \underline{\underline{\Gamma}}_e \Leftrightarrow \underline{\underline{\sigma}}^s = \sum_{k=0}^{2} \hat{G}_k \mathbf{a}_k(\underline{n}) \bullet \underline{\underline{\varepsilon}} + G_3 \mathbf{a}_3(\underline{n}) \bullet \underline{\underline{\Omega}}_e, \;\; \underline{\underline{\sigma}}^a = G_4 \mathbf{a}_4(\underline{n}) \bullet \underline{\underline{\varepsilon}}_e + G_5 \mathbf{a}_5(\underline{n}) \bullet \underline{\underline{\Omega}}_e$$

$$\underline{\underline{\sigma}} = \boldsymbol{\eta}(\underline{n}) \bullet \underline{\underline{\gamma}}_p = \sum_{k=0}^{5} \hat{\eta}_k \mathbf{a}_k(\underline{n}) \bullet \underline{\underline{\gamma}}_p \Leftrightarrow \underline{\underline{\sigma}}^s = \sum_{k=0}^{2} \hat{\eta}_k \mathbf{a}_k(\underline{n}) \bullet \underline{\underline{e}}_p + \eta_3 \mathbf{a}_3(\underline{n}) \bullet \underline{\underline{\omega}}_p, \;\; \underline{\underline{\sigma}}^a = \eta_4 \mathbf{a}_4(\underline{n}) \bullet \underline{\underline{e}}_p + \eta_5 \mathbf{a}_5(\underline{n}) \bullet \underline{\underline{\omega}}_p$$

(2.20$_{1,2}$)

Here $\mathbf{G}(\underline{n}) \equiv \mathbf{R}_G^o(\underline{n})$ and $\boldsymbol{\eta}(\underline{n}) \equiv \mathbf{R}_\eta^o(\underline{n})$ are the ON-operators (or ON-operators) of moduli and viscosity.

2) *N-operator presentations of free energy and dissipation*:

$$f = 1/2\sum_{k=0}^{5} \hat{G}_k \underline{\underline{\Gamma}}_e \bullet \mathbf{a}_k(\underline{n}) \bullet \underline{\underline{\Gamma}}_e; \quad D \equiv TP_s\big|_T = \sum_{k=0}^{5} \hat{\eta}_k \underline{\underline{\gamma}}_p \bullet \mathbf{a}_k(\underline{n}) \bullet \underline{\underline{\gamma}}_p \qquad (2.21_{1,2})$$

3) *Inverse relations expressing kinematic variables via stresses*:

$$\begin{aligned} \underline{\underline{\Gamma}}_e &= \mathbf{G}^{-1}(\underline{n}) \bullet \underline{\underline{\sigma}} \equiv \mathbf{J}(\underline{n}) \bullet \underline{\underline{\sigma}}, \quad \mathbf{J}(\underline{n}) = \sum_{k=0}^{5} J_k \mathbf{a}_k(\underline{n}), \\ \underline{\underline{\gamma}} &= \boldsymbol{\eta}^{-1}(\underline{n}) \bullet \underline{\underline{\sigma}} \equiv \boldsymbol{\Phi}(\underline{n}) \bullet \underline{\underline{\sigma}}, \quad \boldsymbol{\Phi}(\underline{n}) = \sum_{k=0}^{5} \varphi_k \mathbf{a}_k(\underline{n}) \end{aligned} \quad . \qquad (2.22_{1,2})$$

Here $\mathbf{J}(\underline{n}) \equiv \mathbf{N}_J^o(\underline{n})$ and $\boldsymbol{\Phi}(\underline{n}) \equiv \mathbf{N}_\varphi^o(\underline{n})$ are the compliance and fluidity ON-operators, respectively. Their basis scalars, *compliances* $J_k$ (dimensionality of inverse modulus) in ($2.22_1$), and *fluidities* $\varphi_k$ ***(***dimensionality of inverse viscosity) in ($2.22_2$), are:

$$\begin{aligned} &J_0 = \frac{1}{G_0}, \quad J_1 = \frac{(G_3^2 - G_1G_5)/G_0}{G_5(G_0+G_1)-G_3^2}, \quad J_2 = -\frac{3/2(G_1+G_2)/G_0}{3/4G_0+G_1+G_2} \\ &J_3 = -J_4 = \frac{-G_3}{G_5(G_0+G_1)-G_3^2}, \quad J_5 = \frac{G_0+G_1}{G_5(G_0+G_1)-G_3^2}, \end{aligned} \qquad (2.23_1)$$

$$\begin{aligned} &\varphi_0 = \frac{1}{\eta_0}, \quad \varphi_1 = \frac{(\eta_3^2 - \eta_1\eta_5)/\eta_0}{\eta_5(\eta_0+\eta_1)-\eta_3^2}, \quad \varphi_2 = -\frac{3/2(\eta_1+\eta_2)/\eta_0}{3/4\eta_0+\eta_1+\eta_2} \\ &\varphi_3 = \frac{-\eta_3}{\eta_5(\eta_0+\eta_1)-\eta_3^2}, \ \varphi_4 = -\varphi_3, \ \varphi_5 = \frac{\eta_0+\eta_1}{\eta_5(\eta_0+\eta_1)-\eta_3^2} \end{aligned} \qquad (2.23_2)$$

4) *Expressions of* $\underline{\underline{\gamma}}_p = \underline{\underline{e}}_p + \underline{\underline{\omega}}_p$ *via* $\underline{\underline{\Gamma}}_e = \underline{\underline{\varepsilon}}_e + \underline{\underline{\Omega}}_e$ *and inverse* from the dual equations ($2.20_{1,2}$):

$$\underline{\underline{\gamma}}_p = \mathbf{s}(\underline{n}) \bullet \underline{\underline{\Gamma}}_e \ \left( \mathbf{s}(\underline{n}) = \sum_{k=0}^{5} s_k \mathbf{a}_k(\underline{n}) \right); \quad \underline{\underline{\Gamma}}_e = \boldsymbol{\theta}(\underline{n}) \bullet \underline{\underline{\gamma}}_p \ \left( \boldsymbol{\theta}(\underline{n}) = \sum_{k=0}^{5} \theta_k \mathbf{a}_k(\underline{n}) \right) \quad (2.24_{1,2})$$

Here $\mathbf{s}(\underline{n}) = \boldsymbol{\eta}^{-1}(\underline{n}) \bullet \mathbf{G}(\underline{n}) = \boldsymbol{\Phi}(\underline{n}) \bullet \mathbf{G}(\underline{n})$ and $\boldsymbol{\theta}(\underline{n}) = \mathbf{G}^{-1}(\underline{n}) \bullet \boldsymbol{\eta}(\underline{n}) = \mathbf{J}(\underline{n}) \bullet \boldsymbol{\eta}(\underline{n})$ are the N-*operators of relaxation frequencies and relaxation times*, respectively. Using ($2.23_{1,2}$), their basis scalar parameters $s_k$ and $\theta_k$ are calculated as:

$$s_0=\frac{G_0}{\eta_0},\ \ s_1=\frac{\eta_5(G_1\eta_0-G_0\eta_1)+\eta_3(G_0\eta_3-G_3\eta_0)}{\eta_0[\eta_5(\eta_0+\eta_1)-\eta_3^2]},\ \ s_2=\frac{3}{2}\cdot\frac{(G_1+G_2)\eta_0-G_0(\eta_1+\eta_2)}{\eta_0(3/4\eta_0+\eta_1+\eta_2)} \quad (2.25_1)$$

$$s_3=\frac{G_3(\eta_0+\eta_1)-\eta_3(G_0+G_1)}{\eta_5(\eta_0+\eta_1)-\eta_3^2},\ s_4=\frac{G_5\eta_3-G_3\eta_5}{\eta_5(\eta_0+\eta_1)-\eta_3^2},\ \ s_5=\frac{G_5(\eta_0+\eta_1)-G_3\eta_3}{\eta_5(\eta_0+\eta_1)-\eta_3^2}$$

$$\theta_0=\frac{\eta_0}{G_0},\ \ \theta_1=\frac{G_5(\eta_1G_0-\eta_0G_1)+G_3(\eta_0G_3-\eta_3G_0)}{G_0[G_5(G_0+G_1)-G_3^2]},\ \ \theta_2=\frac{3}{2}\cdot\frac{(\eta_1+\eta_2)G_0-\eta_0(G_1+G_2)}{G_0(3/4G_0+G_1+G_2)}$$

$$\theta_3=\frac{\eta_3(G_0+G_1)-G_3(\eta_0+\eta_1)}{G_5(G_0+G_1)-G_3^2},\ \ \theta_4=\frac{\eta_5G_3-\eta_3G_5}{G_5(G_0+G_1)-G_3^2},\ \theta_5=\frac{\eta_5(G_0+G_1)-\eta_3G_3}{G_5(G_0+G_1)-G_3^2} \qquad (2.25_2)$$

Note that $\boldsymbol{\theta}(\underline{n})=\boldsymbol{\eta}(\underline{n})\bullet\mathbf{G}^{-1}(\underline{n})$ and $\mathbf{s}(\underline{n})=\mathbf{G}(\underline{n})\bullet\boldsymbol{\eta}^{-1}(\underline{n})=\boldsymbol{\theta}^{-1}(\underline{n})$ are not ON-operators.

5) *Evolution equation for elastic (transient) strain* $\underline{\underline{\varepsilon}}$ *and elastic rotation* $\underline{\underline{\Omega}}_e$**,** obtained upon substituting (2.24$_1$) in (2.4), is:

$$\dot{\underline{\underline{\Gamma}}}_e+\mathbf{s}(\underline{n})\bullet\underline{\underline{\Gamma}}_e=\underline{\underline{\gamma}}\Leftrightarrow\dot{\underline{\underline{\varepsilon}}}+\sum_{k=0}^{2}s_k\mathbf{a}_k(\underline{n})\bullet\underline{\underline{\varepsilon}}+s_3\mathbf{a}_3(\underline{n})\bullet\underline{\underline{\Omega}}_e=\underline{\underline{e}},\ \ \dot{\underline{\underline{\Omega}}}_e+s_4\mathbf{a}_4(\underline{n})\bullet\underline{\underline{\varepsilon}}+s_5\mathbf{a}_5(\underline{n})\bullet\underline{\underline{\Omega}}_e=\underline{\underline{\omega}} \quad (2.26_1)$$

It is written in the common tensor form is:

$$\begin{aligned}&\dot{\underline{\underline{\varepsilon}}}+s_0\underline{\underline{\varepsilon}}+s_1[\underline{nn}\cdot\underline{\underline{\varepsilon}}+\underline{\underline{\varepsilon}}\cdot\underline{nn}-2nn(\underline{\underline{\varepsilon}}:\underline{nn})]+s_2(\underline{nn}-\underline{\underline{\delta}}/3)(\underline{\underline{\varepsilon}}:\underline{nn})\\&\qquad+s_3(\underline{nn}\cdot\underline{\underline{\Omega}}_e-\underline{\underline{\Omega}}_e\cdot\underline{nn})=\underline{\underline{e}}\\&\dot{\underline{\underline{\Omega}}}_e+s_4(\underline{nn}\cdot\underline{\underline{\varepsilon}}-\underline{\underline{\varepsilon}}\cdot\underline{nn})+s_5(\underline{nn}\cdot\underline{\underline{\Omega}}_e+\underline{\underline{\Omega}}_e\cdot\underline{nn})=\underline{\underline{\omega}}\end{aligned} \qquad (2.26_2)$$

Here the basis scalars of N-operator of relaxation frequency $\mathbf{s}(\underline{n})$ are presented in (2.25$_1$).

6) *Maxwell-like nematodynamic equations* have the following equivalent forms:

$$\mathbf{J}(\underline{n})\bullet\dot{\underline{\underline{\sigma}}}+\boldsymbol{\Phi}(\underline{n})\bullet\underline{\underline{\sigma}}=\underline{\underline{\gamma}},\quad \dot{\underline{\underline{\sigma}}}+\mathbf{s}(\underline{n})\bullet\underline{\underline{\sigma}}=\mathbf{G}(\underline{n})\bullet\underline{\underline{\gamma}}\ ,\ \ \boldsymbol{\theta}(\underline{n})\bullet\dot{\underline{\underline{\sigma}}}+\underline{\underline{\sigma}}=\boldsymbol{\eta}(\underline{n})\bullet\underline{\underline{\gamma}} \qquad (2.27_{1,2,3})$$

The basis scalar parameters $J_k,\varphi_k$ and $s_k,\theta_k$ are expressed via the given model parameters $G_k$ and $\eta_k$ in (2.23$_{1,2}$) and (2.24$_{1,2}$), respectively. The example of "split" expression for CE (2.22$_2$) is:

$$\begin{aligned}&\dot{\underline{\underline{\sigma}}}^s+\sum_{k=0}^{2}\mathbf{a}_k(\underline{n})\bullet(s_k\underline{\underline{\sigma}}^s-G_k\underline{\underline{e}})+\mathbf{a}_3(\underline{n})\bullet(s_3\underline{\underline{\sigma}}^a-G_3\underline{\underline{\omega}})=\underline{\underline{0}}\\&\dot{\underline{\underline{\sigma}}}^a+\mathbf{a}_4(\underline{n})\bullet(s_4\underline{\underline{\sigma}}^s-G_3\underline{\underline{e}})+\mathbf{a}_5(\underline{n})\bullet(s_5\underline{\underline{\sigma}}^a-G_5\underline{\underline{\omega}})=\underline{\underline{0}}\end{aligned} \qquad (2.27_2^*)$$

7) *The eigenvalues of* N-*operator of relaxation frequency* $\mathbf{s}(\underline{n})$ due to (1.18) are:

$$\nu_1=s_0,\ \ \nu_2=s_0+2/3s_2,\ \ \nu_{3,4}=1/2(s_0+s_1+s_5\pm d)\ ,\ \ d^2=(s_0+s_1-s_5)^2-4s_3s_4. \quad (2.28)$$

Here $s_k$ are the basis parameters (2.25$_1$) of the N-operator of relaxation frequency $\mathbf{s}(\underline{n})$, Due to the Theorem 3(iii), all $\nu_k$ in (2.28) are positive, while the N-operator $\mathbf{s}(\underline{n})$ is not necessarily positive.

8) *Basic* ***representation theorem*** *of non-symmetric linear nematic viscoelasticity*:

Maxwell-like nematodynamic equations (2.27) are always presented in the equivalent forms of LEP CE's (2.11) or (2.16), where the parameters are changed for linear viscoelastic functionals.

The N-operator formulation is:

$$\underline{\underline{\sigma}} = \sum_{k=0}^{5} \mathbf{a}_k(\underline{n}) \bullet \{\phi_k(t) * \underline{\underline{\gamma}}(t)\} \quad \left( \phi_k(t) * \underline{\underline{\gamma}}(t) \equiv \int_{-\infty}^{t} \phi_k(t-t_1)\underline{\underline{\gamma}}(t_1)dt_1 \right)$$

$$\underline{\underline{\sigma}}^s = \sum_{k=0}^{2} \mathbf{a}_k(\underline{n}) \bullet (\phi_k * \underline{\underline{e}}) + \mathbf{a}_3(\underline{n}) \bullet (\phi_3 * \underline{\underline{\omega}}), \ \underline{\underline{\sigma}}^a = \mathbf{a}_4(\underline{n}) \bullet (\phi_4 * \underline{\underline{e}}) + \mathbf{a}_5(\underline{n}) \bullet (\phi_5 * \underline{\underline{\omega}}) \tag{2.29$_1$}$$

The common tensor formulation is:

$$\underline{\underline{\sigma}}^s = \phi_0 * \underline{\underline{e}} + \phi_1 * [\underline{nn} \cdot \underline{\underline{e}} + \underline{\underline{e}} \cdot \underline{nn} - 2\underline{nn}(\underline{nn} : \underline{\underline{e}})] + 2(\phi_1 + \phi_2) * (\underline{nn} - \underline{\underline{\delta}}/3)(\underline{nn} : \underline{\underline{e}})$$

$$+ \phi_3 * (\underline{nn} \cdot \underline{\underline{\omega}} - \underline{\underline{\omega}} \cdot \underline{nn}), \ \underline{\underline{\sigma}}^a = -\phi_4 * (\underline{nn} \cdot \underline{\underline{e}} - \underline{\underline{e}} \cdot \underline{nn}) + \phi_5 * (\underline{nn} \cdot \underline{\underline{\omega}} + \underline{\underline{\omega}} \cdot \underline{nn}) \tag{2.29$_2$}$$

Here $\phi_k(t)$ are:

$$\begin{aligned}
&\phi_0(t) = G_0 e^{-\nu_1 t}, \quad \phi_1(t) = -G_0 e^{-\nu_1 t} + (G_0 + G_1)(\kappa_2 e^{-\nu_3 t} - \kappa_1 e^{-\nu_4 t}), \\
&\phi_2(t) = -3/2 G_0 e^{-\nu_1 t} + 2(3/4 G_0 + G_1 + G_2) e^{-\nu_2 t} \\
&\phi_3(t) = -(s_3/d)(G_0 + G_1) e^{-\nu_1 t} + [(s_3/d)(G_0 + G_1) - \kappa_1 G_3 s_4/d] e^{-\nu_3 t} + \kappa_2 G_3 e^{-\nu_4 t} \\
&\phi_4(t) = -(\kappa_2 G_3 + G_5 s_4/d) e^{-\nu_3 t} + (\kappa_1 G_3 + G_5 s_4/d) e^{-\nu_4 t} \\
&\phi_5(t) = -G_3(s_3/d) e^{-\nu_1 t} + [G_3(s_3/d) - \kappa_1 G_5 s_4/d] e^{-\nu_3 t} + \kappa_2 G_5 e^{-\nu_4 t} \\
&\qquad \{\kappa_1 = (s_0 + s_1 - \nu_3)/d, \ \ \kappa_2 = (s_0 + s_1 - \nu_4)/d\}
\end{aligned} \tag{2.30}$$

*Proof*

(i) Consider the first equation in (2.26$_1$), $\dot{\underline{\underline{\Gamma}}}_e + \mathbf{s}(\underline{n}) \bullet \underline{\underline{\Gamma}}_e = \underline{\underline{\gamma}}$. Searching for a solution of the homogeneous equation in the form, $\underline{\underline{\Gamma}}_e^h(t) = \hat{\underline{\underline{\Gamma}}} e^{-\nu t}$, reduces solution of this equation to the eigenvalue problem (1.17$_2$), where $\mathbf{N}_r(\nu, \underline{n}) = \mathbf{s}(\underline{n}) - \nu \mathbf{I}(\underline{n})$ and eigenvalues $\nu_k$ are exposed in (2.28). The eigenoperators $\mathbf{Q}(\nu_k, \underline{n})$ are presented in (1.19), with substitutions $r_k \to s_k$.

(ii) Utilizing additionally the condition (1.21) yields the following solution of initial homogeneous problem: $\underline{\underline{\Gamma}}_e^h(t) = \sum_{k=1}^{4} e^{-\nu_k t} \mathbf{Q}(\nu_k, \underline{n}) \bullet \underline{\underline{\Gamma}}_e^h(0)$. Employing then the standard technique yields the solution of the evolution equation (2.21$_1$) presented in N-operator form as a linear memory functional: $\underline{\underline{\Gamma}}_e(t)=\sum_{k=1}^{4}\mathbf{Q}(\nu_k,\underline{n})\bullet\int_{-\infty}^{t}e^{-\nu_k(t-t_1)}\underline{\underline{\gamma}}(t_1)dt_1$. Using here the relations (1.19) and (1.21) with substitutions $r_k\to s_k$ yields:

$$\underline{\underline{\Gamma}}_e(t)=\sum_{k=0}^{5}\mathbf{a}_k(\underline{n})\bullet\{\chi_k(t)*\underline{\underline{\gamma}}(t)\}\,. \tag{2.31}$$

The following notations have been utilized in (2.31):

$$\begin{aligned}&\chi_0(t)=e^{-\nu_1 t},\ \ \chi_1(t)=-e^{-\nu_1 t}+(\kappa_2/d)e^{-\nu_3 t}-(\kappa_1/d)e^{-\nu_4 t},\ \ \chi_2(t)=3/2(e^{-\nu_2 t}-e^{-\nu_1 t}),\\ &\chi_3(t)=(s_3/d)(e^{-\nu_3 t}-e^{-\nu_1 t}),\ \ \chi_4(t)=(s_4/d)(e^{-\nu_4 t}-e^{-\nu_3 t}),\ \ \chi_5(t)=-\kappa_1 e^{-\nu_3 t}+\kappa_2 e^{-\nu_4 t}\end{aligned} \tag{2.32}$$

(iii) Using (2.31) in the first relation of ($2.16_1$) yields the formulae ($2.29_{1,2}$) and (2.30), which proves the theorem

*Remark*

Formulae (2.29)/(2.30) and (2.31)/(2.32) correctly describe the two limiting cases:

(i) The initial elastic "jump", i.e. $\underline{\underline{\Gamma}}_e(+0)=\underline{\underline{\Gamma}}_0$ and $\underline{\underline{\sigma}}=\mathbf{G}(\underline{n})\bullet\underline{\underline{\Gamma}}_0$, when $\underline{\underline{\gamma}}(t)=\underline{\underline{\Gamma}}_0\delta(t)$, and

(ii) The case: $\underline{\underline{\Gamma}}_e(\infty)=\boldsymbol{\theta}(\underline{n})\bullet\underline{\underline{\gamma}}_0$, $\underline{\underline{\sigma}}(\infty)=\boldsymbol{\eta}(\underline{n})\bullet\underline{\underline{\gamma}}_0$, when $\underline{\underline{\gamma}}(t)=H(t)\underline{\underline{\gamma}}_0$ ($\underline{\underline{\gamma}}_0=const$).

Here $\delta(t)$ and $H(t)$ are Dirac delta and Heaviside functions, and $\mathbf{G}(\underline{n}),\boldsymbol{\theta}(\underline{n})$, and $\boldsymbol{\eta}(\underline{n})$ are the N-operator of moduli, relaxation time, and viscosity, respectively.

### *2.3.2. TI-operator presentation of symmetric theory*

1) *TI-operator presentation of "elastic" and "viscous" CE's* ($2.16_{1,2}$):

$$\underline{\underline{\sigma}}=\mathbf{G}(\underline{n})\bullet\underline{\underline{\varepsilon}}=\boldsymbol{\eta}(\underline{n})\bullet\underline{\underline{e}}_p;\ \ \mathbf{G}(\underline{n})=\sum_{k=0}^{2}\hat{G}_k\mathbf{a}_k(\underline{n}),\ \ \boldsymbol{\eta}(\underline{n})=\sum_{k=0}^{2}\hat{\eta}_k\mathbf{a}_k(\underline{n}) \tag{$2.33_{1,2}$}$$

Here $\mathbf{G}(\underline{n})\equiv\mathbf{S}_G(\underline{n})$ and $\boldsymbol{\eta}(\underline{n})\equiv\mathbf{S}_\eta(\underline{n})$ are the TI operators of moduli and viscosity. Also, $\hat{G}_k$ and $\hat{\eta}_k$ are defined in ($2.19_{1,2}$) for $k=0,1,2$ with no-degenerating conditions $G_k\neq 0$, $\eta_k\neq 0$. It is also worth reminding of simplifying notations, $G_k^r\leftrightarrow G_k$, $\eta_k^r\leftrightarrow\eta_k$ $(k=1,2)$, accepted in Section 2.2.2 after renormalization procedure.

2) *TI-operator presentations of free energy and dissipation*:

$$f=1/2\underline{\underline{\varepsilon}}\bullet\mathbf{G}(\underline{n})\bullet\underline{\underline{\varepsilon}},\quad D=\underline{\underline{e}}_p\bullet\boldsymbol{\eta}(\underline{n})\bullet\underline{\underline{e}}_p\,. \tag{$2.34_{1,2}$}$$

3) *Inverse relations expressing kinematical variables via stresses*:

$$\underline{\underline{\varepsilon}} = \mathbf{G}^{-1}(\underline{n})\bullet\underline{\underline{\sigma}},\ \ \mathbf{G}^{-1}(\underline{n}) \equiv \mathbf{J}(\underline{n}) = \sum_{i=0}^{3} J_k \mathbf{a}_k(\underline{n});\ \ \underline{\underline{\varepsilon}} = \boldsymbol{\eta}^{-1}(\underline{n})\bullet\underline{\underline{\sigma}},\ \ \boldsymbol{\eta}^{-1}(\underline{n}) \equiv \boldsymbol{\Phi}(\underline{n}) = \sum_{i=0}^{3} \varphi_k \mathbf{a}_k(\underline{n}) \qquad (2.35_{1,2})$$

Here $\mathbf{J}(\underline{n})$ and $\boldsymbol{\Phi}(\underline{n})$ are the compliance and fluidity TI-operators, respectively. Due to (1.29) the expressions for their respective basis scalars are:

$$J_0 = 1/G_0,\quad J_1 = -\frac{G_1/G_0}{G_0+G_1},\quad J_2 = -\frac{3/2(G_1+G_2)/G_0}{3/4G_0+G_1+G_2} \qquad (2.36_1)$$

$$\varphi_0 = 1/\eta_0,\quad \varphi_1 = -\frac{\eta_1/\eta_0}{\eta_0+\eta_1},\quad \varphi_2 = -\frac{3/2(\eta_1+\eta_2)/\eta_0}{3/4\eta_0+\eta_1+\eta_2}. \qquad (2.36_2)$$

4) *Expression* $\underline{\underline{e}}_p$ via $\underline{\underline{\varepsilon}}_e$ *or vice versa using dual equations* (2.33$_1$):

$$\underline{\underline{e}}_p = \boldsymbol{\theta}(\underline{n})\bullet\underline{\underline{\varepsilon}},\qquad \underline{\underline{\varepsilon}} = \mathbf{s}(\underline{n})\bullet\underline{\underline{e}}_p\ . \qquad (2.37_{1.2})$$

Here $\mathbf{s}(\underline{n}) = \boldsymbol{\eta}^{-1}(\underline{n})\bullet\mathbf{G}(\underline{n}) = \boldsymbol{\Phi}(\underline{n})\bullet\mathbf{G}(\underline{n})$ and $\boldsymbol{\theta}(\underline{n}) = \mathbf{G}^{-1}(\underline{n})\bullet\boldsymbol{\eta}(\underline{n}) = \mathbf{J}(\underline{n})\bullet\boldsymbol{\eta}(\underline{n})$ are the TI-*operators of relaxation frequencies and relaxation times*, respectively. Using (2.36$_{1,2}$), their basis scalar parameters $s_k$ and $\theta_k$ are calculated as:

$$\theta_0 = \frac{\eta_0}{G_0},\ \ \theta_1 = \frac{\eta_1 G_0 - G_1\eta_0}{G_0(G_0+G_1)},\ \ \theta_2 = \frac{3}{2}\cdot\frac{G_0(\eta_1+\eta_2) - \eta_0(G_1+G_2)}{G_0(3/4G_0+G_1+G_2)} \qquad (2.38_1)$$

$$s_0 = \frac{G_0}{\eta_0},\ \ s_1 = \frac{G_1\eta_0 - \eta_1 G_0}{\eta_0(\eta_0+\eta_1)},\ \ s_2 = \frac{3}{2}\cdot\frac{\eta_0(G_1+G_2) - G_0(\eta_1+\eta_2)}{\eta_0(3/4\eta_0+\eta_1+\eta_2)}. \qquad (2.38_2)$$

5) *Evolution equation for elastic (transient) strain* $\underline{\underline{\varepsilon}}$, obtained upon substituting (2.37$_1$) in (2.4), is:

$$\dot{\underline{\underline{\varepsilon}}} + \mathbf{s}(\underline{n})\bullet\underline{\underline{\varepsilon}} = \underline{\underline{e}} \quad\Leftrightarrow\quad \dot{\underline{\underline{\varepsilon}}} + \sum_{k=0}^{2} s_k \mathbf{a}_k(\underline{n})\bullet\underline{\underline{\varepsilon}} = \underline{\underline{e}}. \qquad (2.39_1)$$

It is written in the common tensor form as:

$$\dot{\underline{\underline{\varepsilon}}} + s_0\underline{\underline{\varepsilon}} + s_1[\underline{n}\underline{n}\cdot\underline{\underline{\varepsilon}} + \underline{\underline{\varepsilon}}\cdot\underline{n}\underline{n} - 2nn(\underline{\underline{\varepsilon}}:\underline{n}\underline{n})] + s_2(\underline{n}\underline{n} - \underline{\underline{\delta}}/3)(\underline{\underline{\varepsilon}}:\underline{n}\underline{n}) = \underline{\underline{e}}\ . \qquad (2.39_2)$$

The basis scalars of TI-operator of relaxation frequency $\mathbf{s}(\underline{n})$ are presented in (2.38$_2$).

6) *Maxwell-like nematodynamic equations* have the following equivalent forms:

$$\mathbf{J}(\underline{n})\bullet\dot{\underline{\underline{\sigma}}} + \boldsymbol{\Phi}(\underline{n})\bullet\underline{\underline{\sigma}} = \underline{\underline{e}},\quad \dot{\underline{\underline{\sigma}}} + \mathbf{s}(\underline{n})\bullet\underline{\underline{\sigma}} = \mathbf{G}(\underline{n})\bullet\underline{\underline{e}},\quad \boldsymbol{\theta}(\underline{n})\bullet\dot{\underline{\underline{\sigma}}} + \underline{\underline{\sigma}} = \boldsymbol{\eta}(\underline{n})\bullet\underline{\underline{e}} \qquad (2.40_{1,2,3})$$

The basis scalar parameters $J_k, \varphi_k$ and $s_k, \theta_k$ are expressed via the given model parameters $G_k$ and $\eta_k$ in (2.36$_{1,2}$) and (2.38$_{1,2}$), respectively.

7) *The eigenvalues of TI-operator of relaxation frequency* $\mathbf{s}(\underline{n})$ due to (1.39) are:

$$\nu_1 = s_0 = \frac{G_0}{\eta_0}, \quad \nu_2 = s_0 + s_1 = \frac{G_0 + G_1}{\eta_0 + \eta_1}, \ \nu_3 = s_0 + \frac{2}{3}s_2 = \frac{3/4G_0 + G_1 + G_2}{3/4\eta_0 + \eta_1 + \eta_2}. \qquad (2.41)$$

Due to the stability conditions ($2.18_{1,2}$) all the eigenvalues $\nu_k$ are positive and describe the *relaxation frequencies*, with the respective *relaxation times* $\hat{\theta}_k = 1/\nu_k$.

8) *Basic* ***representation theorem*** *of symmetric linear nematic viscoelasticity*:

Maxwell-like nematodynamic equations (2.40) are always presented in the equivalent forms of Ericksen CE's ($2.16_2$), where parameters are changed for linear viscoelastic functionals, as:

$$\underline{\underline{\sigma}} = \sum_{k=0}^{2} \mathbf{a}_k(\underline{n}) \bullet \{\phi_k(t) * \underline{\underline{e}}(t)\} \quad \left( \phi_k(t) * \underline{\underline{e}}(t) \equiv \int_{-\infty}^{t} \phi_k(t - t_1)\underline{\underline{e}}(t_1)dt_1 \right) \qquad (2.42_1)$$

or:

$$\underline{\underline{\sigma}} = \phi_0 * \underline{\underline{e}} + \phi_1 * [\underline{nn} \cdot \underline{\underline{e}} + \underline{\underline{e}} \cdot \underline{nn} - 2\underline{nn}(\underline{nn} : \underline{\underline{e}})] + 2(\phi_1 + \phi_2) * (\underline{nn} - \underline{\underline{\delta}}/3)(\underline{nn} : \underline{\underline{e}}) \qquad (2.42_2)$$

With eigenvalues $\nu_k$ given in (2.41), $\phi_k(t)$ are given as:

$$\begin{aligned} &\phi_0(t) = G_0 e^{-\nu_1 t}, \quad \phi_1(t) = G_1 e^{-\nu_2 t} + (G_0 + G_1)(e^{-\nu_2 t} - e^{-\nu_1 t}), \\ &\phi_2(t) = 2(G_1 + G_2)e^{-\nu_1 t} + 2(3/4G_0 + G_1 + G_2)(e^{-\nu_3 t} - e^{-\nu_1 t}) \end{aligned} \qquad (2.43)$$

*Proof*

It is the same as in the non-symmetric case, and based on the solution of evolution equation (2.39) for elastic (transient) strain,

$$\underline{\underline{\varepsilon}} = \sum_{k=0}^{2} \mathbf{a}_k(\underline{n}) \bullet \{\chi_k(t) * \underline{\underline{e}}(t)\} \quad \left[ \chi_0(t) = e^{-\nu_1 t}, \ \chi_1(t) = e^{-\nu_2 t} - e^{-\nu_1 t}, \ \chi_2(t) = \frac{3}{2}(e^{-\nu_3 t} - e^{-\nu_1 t}) \right] (2.44)$$

The same limiting cases as in non-symmetric nematic viscoelasticity are valid here.

### *2.3.3. Soft modes in linear nematic viscoelasticity*

1) Consider now NG TI operators, when the values of material parameters are close to those belonging to the *marginal stability* boundaries in ($2.18_{1,2}$). There are four independent *marginal stability* conditions:

$$G_0 + G_1^r = G_0 + G_1 - G_3^2/G_5 = 0, \quad \eta_0 + \eta_1^r = \eta_0 + \eta_1 - \eta_3^2/\eta_5 = 0 \qquad (2.45_1)$$

$$3/4G_0 + G_1^r + G_2^r = 3/4G_0 + G_1 + G_2 = 0, \ 3/4\eta_0 + \eta_1^r + \eta_2^r = 3/4\eta_0 + \eta_1 + \eta_2 = 0. \qquad (2.45_2)$$

The *nearly marginal*, still stable situations happen when instead ($2.45_{1,2}$) the four independent conditions are satisfied:

$$G_0 + G_1 = \delta_G G_0, \quad \eta_0 + \eta_1 = \delta_\eta \eta_0 \qquad (0 << \delta_G, \delta_\eta << 1) \tag{$2.46_1$}$$

$$3/4G_0 + G_1 + G_2 = 3/2k_G G_0, \quad 3/4\eta_0 + \eta_1 + \eta_2 = 3/2k_\eta \eta_0 \quad (0 << k_G, k_\eta << 1) \tag{$2..46_2$}$$

*If one of (or both) the conditions ($2.46_1$) occurs, the behavior of viscoelastic (as well as the viscous or weakly elastic) nematics is highly sensitive to magnetic field*, with great effects expected when the field is applied. On the contrary, the conditions ($2.46_2$) seem to be insensitive to magnetic field.

2) When the magnetic field is absent, one can asymptotically use the marginal conditions ($2.45_{1,2}$) and employ the results of Section 1.5.4. We consider here the extremal case when both the elastic and viscous TI operators are *completely soft.* Then CE's ($2.33_{1,2}$) take the form:

$$\underline{\underline{\sigma}} = G_0 \boldsymbol{\alpha}(\underline{n}) \bullet \underline{\underline{\varepsilon}} = \eta_0 \boldsymbol{\alpha}(\underline{n}) \bullet \underline{\underline{e}}_p, \quad \boldsymbol{\alpha}(\underline{n}) = \mathbf{a}_0(\underline{n}) - \mathbf{a}_1(\underline{n}) - 3/2\mathbf{a}_2(\underline{n}) \,. \tag{2.47}$$

In this case,

$$G_1 = -G_0, \; G_2 = G_0/4; \quad \eta_1 = -\eta_0, \; \eta_2 = \eta_0/4 \,, \tag{2.48}$$

and free energy and dissipation are represented as:

$$\begin{aligned} f/G_0 &= 1/2\underline{\underline{\varepsilon}} \bullet \boldsymbol{\alpha}(\underline{n}) \bullet \underline{\underline{\varepsilon}} = 1/2\left|\underline{\underline{\varepsilon}}\right|^2 - \underline{n}\underline{n} : \underline{\underline{\varepsilon}}^2 + 1/4(\underline{n}\underline{n} : \underline{\underline{\varepsilon}})^2 \ge 0 \\ 2D/\eta_0 &= 1/2\underline{\underline{e}}_p \bullet \boldsymbol{\alpha}(\underline{n}) \bullet \underline{\underline{e}}_p = 1/2\left|\underline{\underline{e}}_p\right|^2 - \underline{n}\underline{n} : \underline{\underline{e}}_p^2 + 1/4(\underline{n}\underline{n} : \underline{\underline{e}}_p)^2 \ge 0 \end{aligned} \,. \tag{2.49}$$

The TI operator $\boldsymbol{\alpha}(\underline{n})$ is singular, i.e. $\boldsymbol{\alpha}^{-1}(\underline{n})$ does not exist, so do not the formulae (2.36)-(2.37) and generally, (2.38).

3) *Evolution equation for elastic (transient) strain* $\underline{\underline{\varepsilon}}$, for *supersoft case* due to the Theorem 7.2, is represented as:

$$\dot{\underline{\underline{\varepsilon}}} + s_0 \boldsymbol{\alpha}(\underline{n}) \bullet \underline{\underline{\varepsilon}} = \underline{\underline{e}} \quad \Leftrightarrow \quad \dot{\underline{\underline{\varepsilon}}} + s_0[\underline{\underline{\varepsilon}} - \underline{n}\underline{n} \cdot \underline{\underline{\varepsilon}} - \underline{\underline{\varepsilon}} \cdot \underline{n}\underline{n} + 2nn(\underline{\underline{\varepsilon}} : \underline{n}\underline{n}) - 3/2(\underline{n}\underline{n} - \underline{\underline{\delta}}/3)(\underline{\underline{\varepsilon}} : \underline{n}\underline{n})] = \underline{\underline{e}} \tag{2.50}$$

Here the only parameter, $s_0 = G_0/\eta_0$, is the relaxation frequency.

4) *The eigenvalues of supersoft TI-operator of relaxation frequency* $\mathbf{s}(\underline{n}) = s_0 \boldsymbol{\alpha}(\underline{n})$ due to (1.35) are:

$$\nu_1 = s_0 = G_0 / \eta_0, \quad \nu_2 = \nu_3 = 0. \tag{2.51}$$

5) The *basic* ***representation theorem*** *of symmetric linear nematic viscoelasticity in the supersoft case* is presented by the limit singular case of ($2.42_1$) as:

$$\underline{\underline{\sigma}} = G_0 \boldsymbol{\alpha}(\underline{n}) \bullet \underline{\underline{E}}(t), \qquad \underline{\underline{E}}(t) = \int_{-\infty}^{t} \underline{\underline{e}}(t_1) e^{-s_0(t-t_1)} dt_1$$

$$\underline{\underline{\sigma}} = G_0[\underline{\underline{E}} - \underline{n}\underline{n} \cdot \underline{\underline{E}} - \underline{\underline{E}} \cdot \underline{n}\underline{n} + 2nn(\underline{\underline{E}} : \underline{n}\underline{n}) - 3/2(\underline{n}\underline{n} - \underline{\underline{\delta}}/3)(\underline{\underline{E}} : \underline{n}\underline{n})] \tag{2.52}$$

*Proof* is based on direct use of (2.51) and the marginal stability conditions (2.45).

*Remark*

The form of convolution formula (2.44) for elastic (transient) strains $\underline{\underline{\varepsilon}}$ remains the same, but the functions $\chi_k(t)$ there change for:

$$\chi_0(t) = e^{-s_0 t}, \ \chi_1(t) = 1 - e^{-s_0 t}, \ \chi_2(t) = 3/2(1 - e^{-s_0 t}). \tag{2.53}$$

It means that the elastic strain tensor $\underline{\underline{\varepsilon}}$ can *unrestrictedly grow* in time. This is the asymptotic effect of the supersoft nematic viscoelastic behavior. Nevertheless, one should be reminded that the linear nematic viscoelasticity can hold only either for a restricted time with a given constant value of strain rate $\underline{\underline{e}}$, or for a small amplitude oscillatory flow.